\documentclass[twocolumn,showpacs,preprintnumbers,amsmath,amssymb,superscriptaddress]{revtex4-2}

\usepackage{epsf}
\usepackage{graphicx}
\usepackage{sidecap}

 \usepackage{array}
 \usepackage{amsmath}
 \usepackage{amssymb}
 \usepackage{dcolumn}
 \usepackage{epstopdf}
 \usepackage{bm}% bold math
 \usepackage{amsmath}

\usepackage{color}
\usepackage{hyperref}
\usepackage{soul}
\sethlcolor{green}
\hypersetup{
    colorlinks=true,
    citecolor=red,
    linkcolor=blue,
    filecolor=blue,   
    urlcolor=magenta,
}
 
\def \beq {\begin{equation}}
\def \eeq {\end{equation}}
\begin{document}
 
\title{{Observation of multiple nodal-lines in $\mathrm{\textbf{SmSbTe}}$ }}
 \author{Sabin~Regmi}\thanks{These authors contributed equally to this work.}\affiliation {Department of Physics, University of Central Florida, Orlando, Florida 32816, USA}
\author{Gyanendra~Dhakal}\thanks{These authors contributed equally to this work.}\affiliation {Department of Physics, University of Central Florida, Orlando, Florida 32816, USA}
\author{Fairoja~Cheenicode~Kabeer}\affiliation{Department of Physics and Astronomy, Uppsala University, P. O. Box 516, S-75120 Uppsala, Sweden}
\author{Neil Harrison} \affiliation {National High Magnetic Field Laboratory, Los Alamos, New Mexico 87545, USA}
\author{Firoza~Kabir}\affiliation {Department of Physics, University of Central Florida, Orlando, Florida 32816, USA} 
\author{Anup~Pradhan~Sakhya}\affiliation {Department of Physics, University of Central Florida, Orlando, Florida 32816, USA}
\author{Krzysztof~Gofryk} \affiliation{Idaho National Laboratory, Idaho Falls, Idaho 83415, USA}
\author{Dariusz~Kaczorowski}\affiliation{Institute of Low Temperature and Structure Research,Polish Academy of Sciences, 50-950 Wroclaw, Poland}
\author{Peter~M.~Oppeneer}\affiliation{Department of Physics and Astronomy, Uppsala University, P. O. Box 516, S-75120 Uppsala, Sweden}
\author{Madhab~Neupane}\thanks{Corresponding author: madhab.neupane@ucf.edu}\affiliation {Department of Physics, University of Central Florida, Orlando, Florida 32816, USA}
 
\date{\today}
 
\begin{abstract}
{
Having been a ground for various topological fermionic phases, the family of ZrSiS-type 111 materials has been under experimental and theoretical investigations. Within this family of materials, the subfamily  $\mathrm{\textit{Ln}SbTe}$ ($Ln = \mathrm{lanthanide~elements}$) is gaining interests in  recent times as the strong correlation effects and magnetism arising from the 4\textit{f} electrons of the lanthanides can provide an important platform  to study the linking between topology, magnetism, and correlation. In this paper, we report the systematic study of the electronic structure of $\mathrm{SmSbTe}$ - a member of the $\mathrm{\textit{Ln}SbTe}$ subfamily - by utilizing angle-resolved photoemission spectroscopy in conjunction with first-principles calculations, transport, and magnetic measurements. Our experimental results identify multiple Dirac nodes forming the nodal-lines along the $\Gamma-X$ and $Z-R$ directions in the bulk Brillouin zone (BZ) as predicted by our theoretical calculations. A surface Dirac-like state that arises from the square net plane of the Sb atoms is also observed at the $\overline{X}$ point of the surface BZ. Our study highlights $\mathrm{SmSbTe}$ as a promising candidate to understand the topological electronic structure of $\mathrm{\textit{Ln}SbTe}$ materials. % The presence of topology and the low temperature magnetism point to $\mathrm{SmSbTe}$ being a promising material candidate to study the linking of magnetism with topology in this highly renowned family of ZrSiS-type materials. 
}
\end{abstract}
\maketitle

\begin{center} \textbf{I. INTRODUCTION} \end{center}
The field of topological quantum materials has been growing ever since the discovery of the three-dimensional (3D) topological insulators  \cite{ti1, ti2, ti3, ti4}. Following the inflow of theoretical and experimental research studies in the field, topological semimetals including the Dirac semimetals  \cite{tsm1, dsm2}, Weyl semimetals  \cite{wsm1, wsm2}, nodal-line/loop semimetals (NLSMs) \cite{tsm3, zrsis1, zrsis2} and beyond \cite{bradlyn, CoSi} were discovered. After the discovery of the nodal-line topological state in ZrSiS \cite{zrsis1, zrsis2}, the ZrSiS-type 111 materials have attracted a lot of research interests. The materials in this family are shown to host nonsymmorphic topological fermions coming from the square net plane of the Group-IV elements (Si, Ge, Sb) and nodal-line fermions \cite{zrsis1, zrsis2, topp, hu, takane, zrsix, chen, lou, hu2, zhang, zrgete, fu}.  The existence of exotic phenomena like unconventional magneto-transport behavior \cite{ali, lv, singha, kumar}, flat optical conductivity \cite{schilling}, unconventional mass enhancement \cite{pezzini}, etc. reported in this family of materials enticed studies of more members of this family of materials.

Among various materials in this family, the materials under the subfamily $\mathrm{\textit{Ln}SbTe} ~[ \textit{Ln} = \mathrm{lanthanide~elements}] $ are of particular interest because of the potential co-existence of topology with magnetic ordering carried by the lanthanides. Furthermore, the lanthanide elements come with strongly correlated 4\textit{f} electrons, which can potentially give a way to study the interplay between topology, correlation, and magnetism in the ZrSiS-type materials. This makes the study of the detailed electronic structure of $\mathrm{\textit{Ln}SbTe}$-type materials desirable, however, only a few studies have been carried out to date. $\mathrm{GdSbTe}$ is reported to exhibit a topological nodal-line state as well as antiferromagnetic Dirac state protected by the combination of broken time-reversal symmetry and rotoinversion symmetry by using angle-resolved photoemission spectroscopy (ARPES) \cite{gdsbte}. Because of the tunability of the magnetic ordering of the Ce 4\textit{f} electrons, $\mathrm{CeSbTe}$ is reported to host a variety of topological features in the electronic structure \cite{cesbte}. Another study on  $\mathrm{CeSbTe}$ showed that the stronger spin-orbit coupling (SOC) in this material creates a more symmetric Dirac cone which is protected by the nonsymmorphic symmetry \cite{cesbte2}. The nodal-line state in $\mathrm{HoSbTe}$ is gapped out in the order of 100s of meV which can be directly observed via ARPES \cite{hosbte}. $\mathrm{NdSbTe}$ has been shown to exhibit coexistence of metamagnetic transitions and possible Kondo localization \cite{ndsbte}, while the Kondo mechanism has been reported in $\mathrm{CeSbTe}$ \cite{cesbte3}. $\mathrm{LaSbTe}$ has been identified as a genuine NLSM with the nodal line state present even with the inclusion of SOC effect \cite{lasbte}. Despite all these works on $\mathrm{\textit{Ln}SbTe}$, $\mathrm{Sm}$-variant of this family has not been studied yet.

\begin{figure*} [ht!]
\includegraphics[width=1\textwidth]{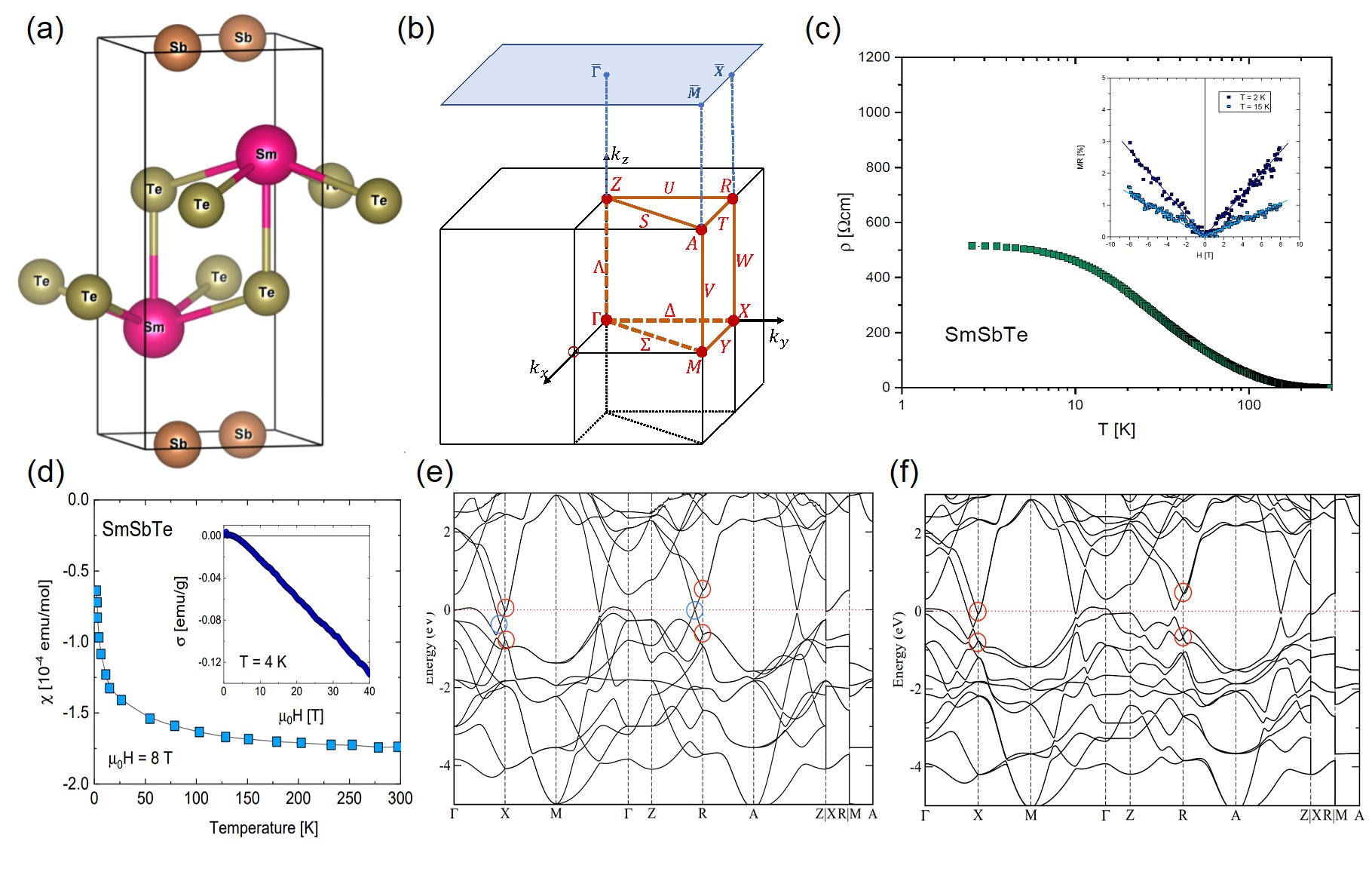}
\caption{Crystal structure and sample characterization of $\mathrm{SmSbTe}$. (a) Unit cell of the $\mathrm{SmSbTe}$ crystal structure. (b) Bulk Brillouin zone and the projected surface Brillouin zone with the high-symmetry points marked. (c) The temperature dependence of electrical resistivity of the  $\mathrm{SmSbTe}$ crystal. The inset shows the magnetic field dependence of the magnetoresistance. The solid lines are a guide for an eye. (d) Temperature dependence of magnetic susceptibility measured in 8 T. The inset shows the field dependence of magnetization measured up to 40 T. (e)-(f) Calculated bulk band structures  without  and  with the consideration of the effects of SOC, respectively.}
\end{figure*}
In this paper, we report the systematic ARPES study of $\mathrm{SmSbTe}$ with parallel first-principles calculations, transport, and magnetic measurements in order to uncover the underlying electronic structure, topology, and magnetism in this compound. The low temperature electronic structure reported in our study shows the presence of typical $\mathrm{ZrSiS}$-type diamond shaped Fermi surface formed by the nodal-line states. Our experimental observations reveal multiple Dirac nodes along bulk $\Gamma-X$ and $Z-R$ directions that form the nodal-lines in concurrence with the theoretical calculations. Furthermore, the experimental results reveal the presence of a Dirac-like state around the $\overline{X}$ point of the surface Brillouin zone (BZ), which comes from the Sb-square nets.  Our study provides a new variant in order to understand the underlying low temperature electronic structure in the $\mathrm{\textit{Ln}SbTe}$ materials. \\
 %Our transport measurements show that $\mathrm{SmSbTe}$ is magnetic at low temperatures. Presence of topology and magnetism makes this material a promising candidate to explore the interplay between magnetism and topology in the ZrSiS-type family of materials. \\

\begin{figure*} [ht!]
\includegraphics[width=1\textwidth]{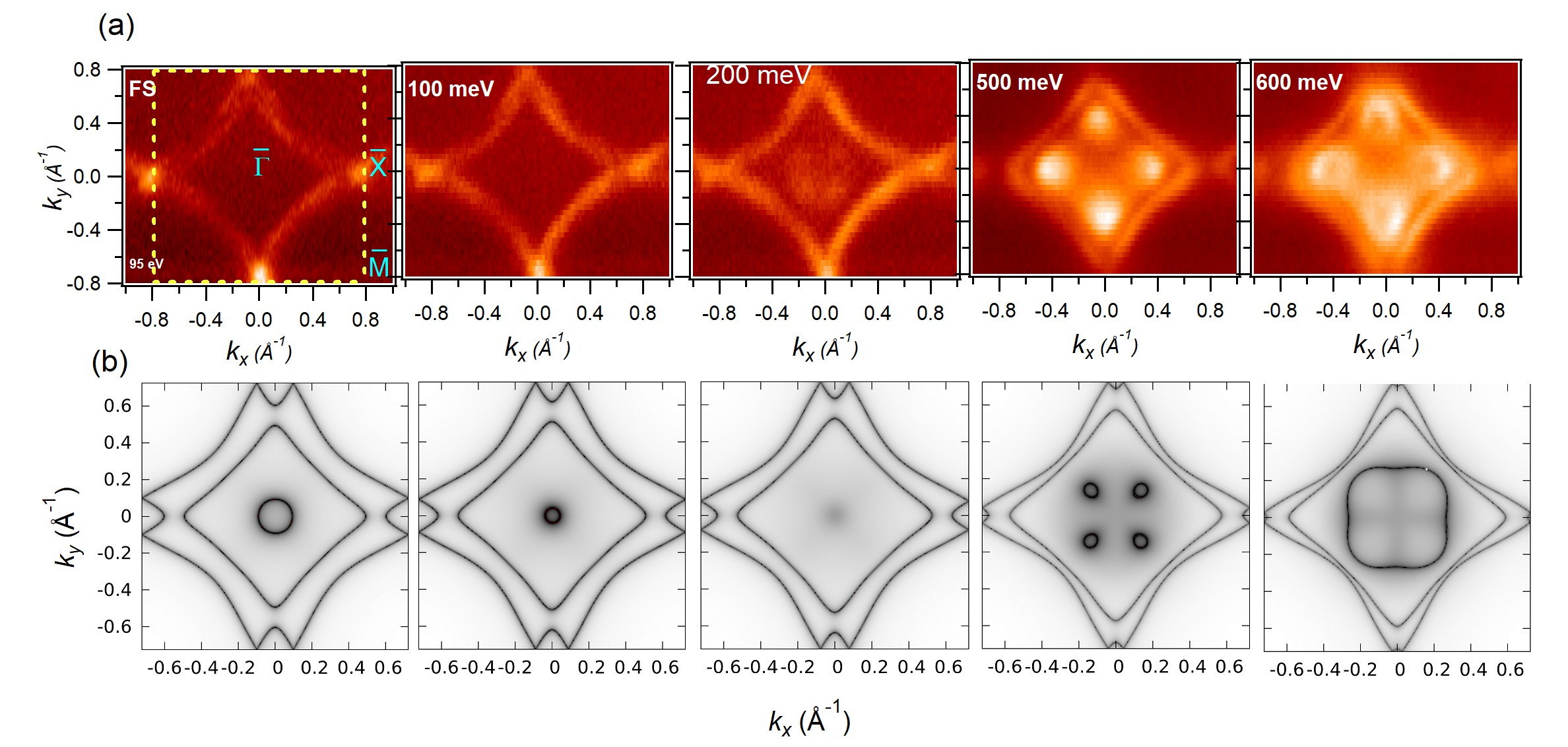}
\caption{Fermi map and constant energy contours. (a) ARPES measured Fermi surface (top leftmost) and constant energy contours on the (001) surface of $\mathrm{SmSbTe}$. The yellow-dashed square on the FS plot represents the surface Brillouin zone. Corresponding binding energies are marked on each of the constant energy contour plots. (b) Calculated Fermi map and constant energy contours corresponding to (a). ARPES data were collected at the SIS-HRPES end station at the SLS, PSI with a photon energy of 95 eV and at a temperature of 20 K.  }
\end{figure*}

\begin{center} \textbf{II. METHODS} \end{center}
Single crystals of $\mathrm{SmSbTe}$ were grown using flux method and characterized as explained in the supplemental material (SM) \cite{SI}. Electrical resistance, magnetoresistance, and magnetic susceptibility (VSM) were measured from $2 - 300 \mathrm{K}$ in magnetic fields up to $14 \mathrm{T}$ using a Quantum Design DynaCool-14 System. Magnetization measurements in pulsed magnetic fields up to $40 \mathrm{T}$ were measured using a pickup-coil technique at NHMFL, Los Alamos. Synchrotron-based ARPES measurements were performed at the SLS SIS-X09LA BL at a temperature of $20 \mathrm{K}$. The first-principles calculations were performed within the density-functional theory (DFT) formalism  \cite{Hohe64,Kohn65} on the basis of projector augmented wave potential \cite{Bloo94} using the Vienna ab initio simulation package (VASP) \cite{Kres96,Kres96.1}. For more details, see Section I of the SM \cite{SI}.

\begin{center} \textbf{III. RESULTS} \end{center}
\begin{center} \textbf{A. Crystal structure, sample characterization, and bulk band calculations} \end{center}
$\mathrm{SmSbTe}$ is a $\mathrm{PbFCl}$-type telleride that crystallizes in a tetragonal layered crystal structure with P4/\textit{nmm} (\#129) nonsymmorphic space group. The $\mathrm{Sb}$ square plane is sandwiched by the $\mathrm{Sm-Te}$ layers forming the quintuple layers of  $\mathrm{Te-Sm-Sb-Sm-Te}$ slabs along the [001] directions. Each quintuple layer is weakly bonded with the neighboring layers via van der Waals interactions, therefore the crystals usually cleave at the $\mathrm{Te}$ termination on the (001) plane. Each layer of atoms in this crystal supports global $C_{4\nu}$ symmetry. Since the Sb layer act as a glide mirror plane, it breaks the $C_{4\nu}$ symmetry locally at the Sb atom sites, which is in line with the   $\left( M_z|\frac{1}{2}\frac{1}{2}0\right)$ symmetry operation.   A unit cell for $\mathrm{SmSbTe}$ crystal structure is presented in Fig. 1(a). The theoretically optimized lattice constants are $\mathrm{a = b = 4.338 ~ \AA ~ and ~ c = 9.398 ~ \AA}$. The 3D bulk BZ  is presented in Fig. 1(b) where the high-symmetry points are indicated. The bulk BZ is projected onto the (001) surface on which the ARPES measurements are performed.  The electrical resistivity of the $\mathrm{SmSbTe}$ crystals is plotted against the temperature in Fig. 1(c). Interestingly, the  resistivity of $\mathrm{SmSbTe}$ strongly increases upon cooling and then saturates at $\sim$8 K. It has been shown that in many systems with surface Dirac states, the saturation of the resistance occurs below temperatures where the metallic topological surface state resistance starts to dominate over the strongly increasing bulk resistance, even for bulk crystals \cite{Ando1, Ando2}. The magnetoresistance (MR) plotted in the inset of Fig. 1(c) shows linear field dependency, which signals towards the possible linear dispersion in the low-temperature electronic structure. The magnetic susceptibility presented in Fig. 1(d) shows typical characteristics of weak diamagnetism. The $\chi(T)$ curve is negative and nearly temperature-independent and the magnetization measured in magnetic fields up to 45 T shows a negative slope of M(H), typical for diamagnetic materials (see the inset of Fig. 1d).  The calculated bulk band structures without and with the inclusion of SOC are shown in Figs. 1 (e) and (f), respectively. In the presence of SOC, the Sm \textit{d} and Sb \textit{p} bands along $Z - A$ direction couple together resulting into band inversion and gap opening indicating the non-trivial topology in $\mathrm{SmSbTe}$ (see SM \cite{SI}). The calculations suggest that in the absence of SOC, multiple Dirac nodal-lines are present along different high-symmetry directions as indicated by red and blue circles in Fig. 1(e).  Some of these Dirac nodal-lines are gapped out when the effect of SOC is taken into account with significant gap size especially along the $Z - R$ direction (see Fig. 1(f)). A similar SOC gap opening was recently reported in $\mathrm{HoSbTe}$ \cite{hosbte}.  On the other hand, some Dirac nodal-lines remain robust even in the presence of SOC (see red circles in Fig. 1(f)).

\begin{figure*} [ht!]
\includegraphics[width=1\textwidth]{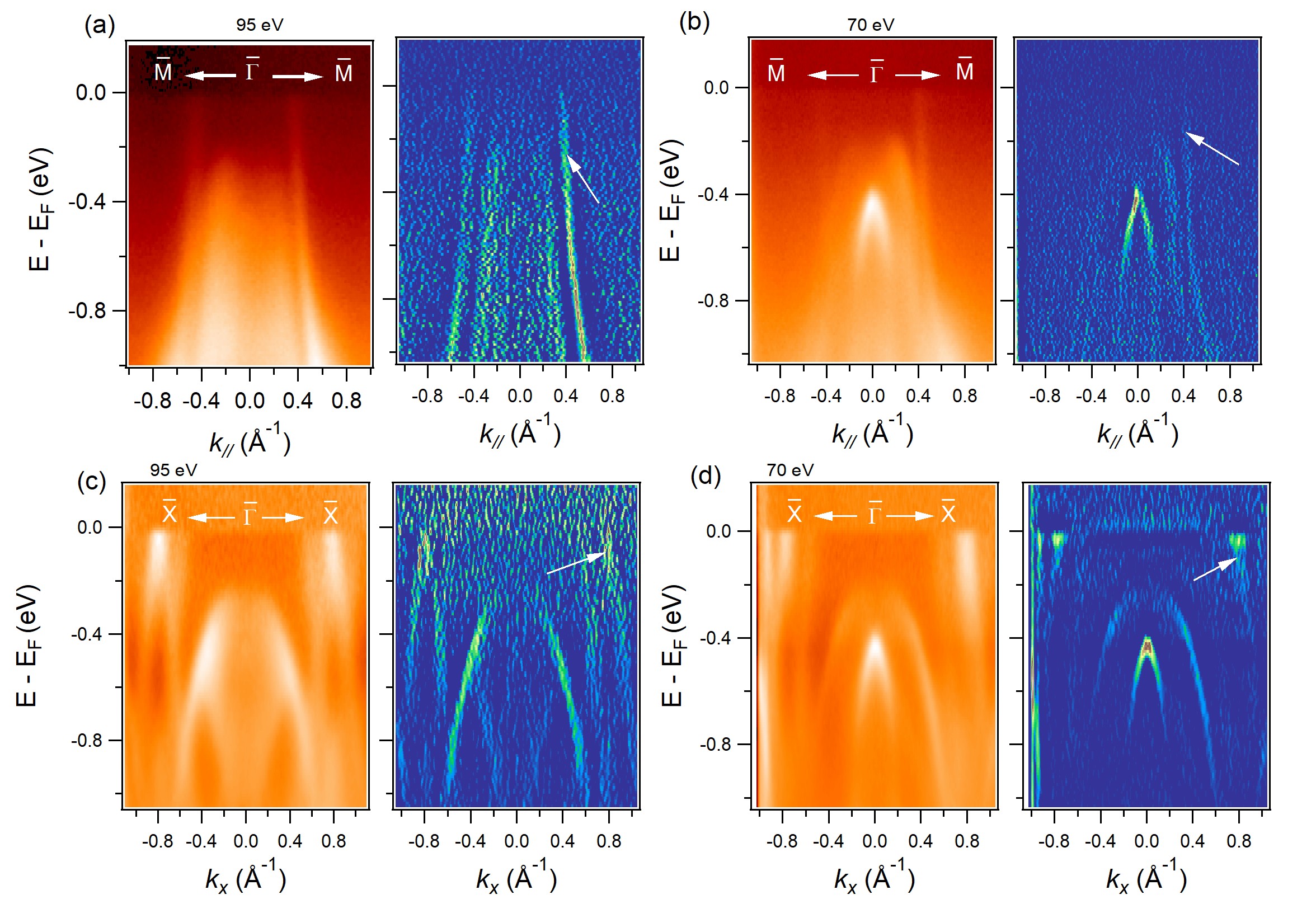}
\caption{ Observation of nodal-line semimetal in $\mathrm{SmSbTe}$. (a) Dispersion map and its second derivatives  along the $\overline{M} - \overline{\Gamma} - \overline{M}$ direction measured at $k_z$ = 0. (b) Dispersion map and its second derivatives  along the $\overline{M} - \overline{\Gamma} - \overline{M}$ direction measured at $k_z$ = $\pi$. (c) Dispersion map and its second derivatives  along the $\overline{X} - \overline{\Gamma} - \overline{X}$ direction measured at $k_z$ = 0. (d) Dispersion map and its second derivatives  along the $\overline{M} - \overline{\Gamma} - \overline{M}$ direction measured at $k_z$ = $\pi$.  ARPES data were collected at the SIS-HRPES end station at the SLS, PSI with a photon energy of 95 eV and at a temperature of 20 K. }
\end{figure*}

\begin{center} \textbf{B. ARPES measured energy contours} \end{center}
In order to investigate the detailed electronic structure of SmSbTe,  we present the experimentally obtained constant energy contours at the Fermi energy (top leftmost plot) and at various binding energies measured with a photon energy of 95 eV, which corresponds to $k_z$=0 (see SI for $k_z$ dependent measurements \cite{SI}). The Fermi surface is a typical of ZrSiS-type materials with a diamond shaped Fermi pocket centering the $\overline{\Gamma}$ point. At around 200 meV below the Fermi level, a faint circular feature begins to appear around the $\overline{\Gamma}$ point. This circular feature grows in its size on moving to higher binding energies and emerges into a second diamond shaped energy pocket at around 500 meV below the Fermi level. The corners of this second diamond shaped feature are surrounded by small circular pockets at this binding energy. These circular pockets evolve with binding energy and become clear ring-like pockets around 600 meV below the Fermi level. Figure 2(b) shows the calculated Fermi map and constant energy contours corresponding to Fig. 2(a). The calculated constant energy contours match well with the experimental plots. The features seen at the $\overline{\Gamma}$ point near the Fermi level in calculated maps are absent in the experimental plots, which are probably due to matrix element effects. Even though the diamond shaped Fermi surface looks single sheet in this photon energy which appears similar to other two $Ln$SbTe compounds \cite{gdsbte, lasbte}, the Fermi map obtained at low photon energy (35 eV) clearly depicts double sheet diamond shape \cite{SI}.

\begin{figure*} [ht!]
\includegraphics[width=1\textwidth]{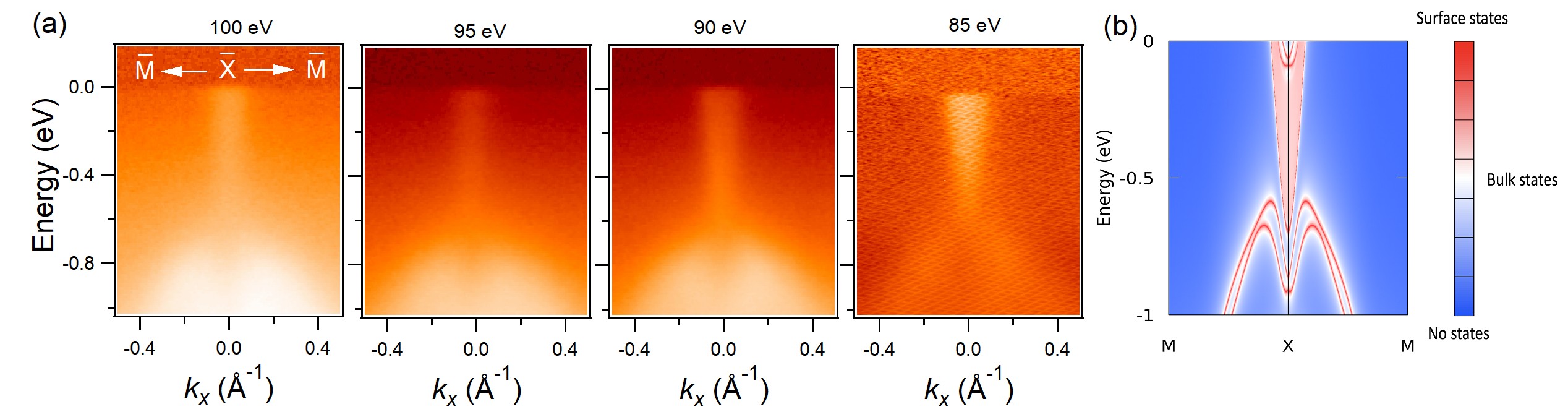}
\caption{Observation of surface state on the (001) surface of $\mathrm{SmSbTe}$. (a) Dispersion maps along the $\overline{M} - \overline{X} - \overline{M}$ direction measured at different photon energies. The corresponding photon energies are noted on top of each plots.  (b) Calculated surface electronic structure along the  $\overline{M} - \overline{X} - \overline{M}$.  ARPES data were collected at the SIS-HRPES end station at the SLS, PSI with a photon energy of 95 eV and at a temperature of 20 K. }
\end{figure*}

\begin{center} \textbf{C. Observation of multiple nodal-lines} \end{center}
In Fig. 3, we present the ARPES measured dispersion maps along the $\overline{M}-\overline{\Gamma}-\overline{M}$ and $\overline{X}-\overline{\Gamma}-\overline{X}$ directions, respectively at two different $k_z$ planes. Figure 3(a) displays the dispersion map along the $\overline{M}-\overline{\Gamma}-\overline{M}$ direction at $k_z$=0. A linearly dispersing band seems to cross the Fermi level, which gives rise to the diamond shaped Fermi pocket. Even though bulk band calculations suggest a gapped state without and with SOC along this direction, we can not resolve the gap in our experimental data probably because of the gap being beyond our experimental resolution. The separation of the bands forming the gapped state may be very close, therefore they seem to be a single band over a large energy range ($\geq$ 1 eV) in the vicinity of Fermi level. %Even though we expect a pair of linear bands along the $\Gamma-M$ direction from bulk band calculations, one can see a single band over a large energy range ($\geq$ 1 eV) in the vicinity of Fermi level. Such observations are reported in previously studied some nodal-line semimetals, which can  possibly be attributed to matrix element effects \cite{gdsbte, lasbte}.
Next, we present dispersion map along the $\overline{M}-\overline{\Gamma}-\overline{M}$ direction at $k_z$=$\pi$ in Fig. 3(b). The bands along this direction show significant changes with photon energies indicating the bulk nature of the bands. Figure 3(c) presents dispersion map ($k_z$=0) along the $\overline{X}-\overline{\Gamma}-\overline{X}$ direction which depicts linearly dispersing bands in the vicinity of $\overline{X}$. Those bands form a nodal-line as predicted by bulk-band calculations. We can see that the Dirac-like crossing lies very close to the Fermi level. The hole-like bands away from the Fermi level around the $\overline{\Gamma}$ make a circular feature at constant energy contours as seen in Fig. 2(a). In Fig. 3(d), the dispersion map along the $\overline{X}-\overline{\Gamma}-\overline{X}$ ($k_z$=$\pi$) is presented, which shows the presence of Dirac-like bands forming the nodal-line state in the bulk calculation along $Z-R$ direction (Fig. 1(f)).  The dispersion maps along these two different directions at two photon energies provide the indication of the bulk bands at the $\overline{\Gamma}$ point, whereas, we can clearly see the coexistence of surface and bulk bands in the vicinity of $\overline{X}$.
 
Next, we focus on the dispersion map along the $\overline{M} - \overline{X} - \overline{M}$ direction which shows the Dirac-like dispersion.  The linear dispersions at the edge of the BZ originates from the $Sb$ square-net. The linear bands do not disperse with the photon energies indicating that the bands correspond to the surface (see SM \cite{SI} for more photon energy dependent measurements). The Dirac node exists around 600 meV below the Fermi level. Figure 4(b) displays the surface calculations along the $\overline{M} - \overline{X} - \overline{M}$ direction, which exhibits small gap at the Dirac node. This gap is not resolved in our experimental measurements, likely due to the experimental resolution being greater than the gap size.  \\

\begin{center} \textbf{III. CONCLUSION} \end{center}
In summary, we performed a detailed ARPES study on $\mathrm{SmSbTe}$ together with transport and magnetic measurements as well as first-principles calculations. Our ARPES data show the typical $\mathrm{ZrSiS}$-type diamond Fermi surface and Dirac-like states at the corner of the BZ. The surface Dirac-like state at the $\overline{X}$ point in the surface BZ is gapped according to our theoretical calculations, however the gap is too small to be resolved from our experiments. Multiple Dirac nodes  forming the nodal-line states are present along  different high-symmetry directions in the absence of SOC, some of which are gapped out and the others remain robust in the presence of SOC.  In addition,  we observe the saturation of the electrical resistivity at low temperatures and linear magnetoresistance through our transport measurements. Our study provides a new platform in order to understand the electronic structure in $\mathrm{\textit{Ln}SbTe}$ subfamily of materials.\\ \\

\begin{center} \textbf{ACKNOWLEDGEMENTS} \end{center}
M.N. is supported by the Air Force Office of Scientific Research under award number FA9550-17-1-0415 and the National Science Foundation (NSF) CAREER award DMR-1847962. P.M.O. acknowledges support from the Swedish Research Council (VR) and the Knut and Alice Wallenberg Foundation (Grant No. 2015.0060). Computational resources were provided by the Swedish National Infrastructure for Computing (SNIC) (Grant No. 2018-05973). K.G. acknowledges support from the INL Laboratory Directed Research and Development (LDRD) Program under DOE Idaho Operations Office Contract DE-AC07-05ID14517. N.H. acknowledges support from the US DOE Basic Energy Science program through the project ''Science at 100T" at LANL. A portion of this work was performed at the National High Magnetic Field Laboratory, which is supported by the National Science Foundation Cooperative Agreement No. DMR-1644779 and the state of Florida. D.K. was supported by the National Science Centre (Poland) under research Grant No. 2015/18/A/ST3/00057. We are grateful to Nicholas Clark Plumb, Ming Shi, Hang Li, and Sailong Ju for beamline assistance at PSI, SLS.
\\


\begin{thebibliography}{50}
	
\bibitem{ti1} M. Z. Hasan and C. L. Kane, Colloquium: Topological insulators, \href{https://doi.org/10.1103/RevModPhys.82.3045}{Rev. Mod. Phys. \textbf{82,} 3045 (2010)}.

\bibitem{ti2}  X.-L. Qi and S.-C. Zhang,  Topological insulators and superconductors,  \href{https://doi.org/10.1103/RevModPhys.83.1057}{Rev. Mod. Phys. \textbf{83,} 1057 (2011)}.

\bibitem{ti3} M. Z. Hasan, S.-Y. Xu, and M. Neupane, in \textit{Topological Insulators: Fundamentals and Perspectives}, edited by S. R. F. Ortmann, and S. O. Valenzuela (Wiley, Hoboken, NJ, 2015).

\bibitem{ti4} A. Bansil, H. Lin,  and T. Das,  Colloquium: Topological band theory,  \href{https://doi.org/10.1103/RevModPhys.88.021004}{Rev. Mod. Phys. \textbf{88,} 021004 (2016)}.

\bibitem{tsm1} N. P. Armitage, E. J. Mele,  and A. Vishwanath,  Weyl and Dirac semimetals in three-dimensional solids,  \href{https://doi.org/10.1103/RevModPhys.90.015001}{Rev. Mod. Phys. \textbf{90,} 015001 (2018)}.

%\bibitem{dsm1} Z. K. Liu, B. Zhou, Y. Zhang, Z. J. Wang, H. M. Weng, D. Prabhakaran, S. K. Mo, Z. X. Shen, Z. Fang, X. Dai, Z. Hussain, and Y. L. Chen, Discovery of a Three-Dimensional Topological Dirac Semimetal ${\mathrm{Na}_3\mathrm{Bi}}$,  \href{https://doi.org/10.1126/science.1245085}{Science \textbf{343,} 864 (2014)}.

\bibitem{dsm2} M. Neupane, S.-Y. Xu, R. Sankar, N. Alidoust, G. Bian, C. Liu, I. Belopolski, T.-R. Chang, H.-T. Jeng, H. Lin, A. Bansil, F. Chou, and M. Z. Hasan,  Observation of a three-dimensional topological Dirac semimetal phase in high-mobility ${\mathrm{Cd}}_3{\mathrm{As}}_2$,  \href{https://doi.org/10.1038/ncomms4786}{Nat. Commun. \textbf{5,} 3786 (2014)}.

%\bibitem{tsm2} B. Yan and C. Felser, Topological Materials: Weyl Semimetals, \href{https://doi.org/10.1146/annurev-conmatphys-031016-025458}{Annu. Rev. Condens. Matter Phys. \textbf{8,} 337 (2017)}.

\bibitem{wsm1} Xu, S. Y. et al. S.-Y. Xu, I. Belopolski, N. Alidoust, M. Neupane, G. Bian, C. Zhang, R. Sankar, G. Chang, Z. Yuan, C.-C. Lee, S.-M. Huang, H. Zheng, J. Ma, D. S. Sanchez, B. Wang, A. Bansil, F. Chou, P. P. Shibayev, H. Lin, S. Jia, and M. Z. Hasan, Discovery of a Weyl fermion semimetal and topological Fermi arcs, \href{https://doi.org/10.1126/science.aaa9297} {Science \textbf{349,} 613 (2015)}.

\bibitem{wsm2} B. Q. Lv, H. M. Weng, B. B. Fu, X. P. Wang, H. Miao, J. Ma, P. Richard, X. C. Huang, L. X. Zhao, G. F. Chen, Z. Fang, X. Dai, T. Qian, and H. Ding, Experimental Discovery of Weyl Semimetal  $\mathrm{TaAs}$, \href{https://doi.org/10.1103/PhysRevX.5.031013} {Phys. Rev. X \textbf{5,} 031013 (2015)}.

\bibitem{tsm3} A. A. Burkov, M. D. Hook, and L. Balents, Topological nodal semimetals, \href{https://doi.org/10.1103/PhysRevB.84.235126}{Phys. Rev. B \textbf{84,} 235126 (2011)}.

\bibitem{zrsis1} L. M. Schoop, M. N. Ali, C. Straßer, A. Topp, A. Varykhalov, D. Marchenko, V. Duppel, S. S. P. Parkin, B. V. Lotsch, and C. R. Ast, Dirac cone protected by non-symmorphic symmetry and three-dimensional Dirac line node in $\mathrm{ZrSiS}$,  \href{https://doi.org/10.1038/ncomms11696}{Nat. Commun. \textbf{7,} 11696 (2016)}.

\bibitem{zrsis2} M. Neupane, I. Belopolski, M. M. Hosen, D. S. Sanchez, R. Sankar, M. Szlawska, S.-Y. Xu, K. Dimitri, N. Dhakal, P. Maldonado, P. M. Oppeneer, D. Kaczorowski, F. Chou, M. Z. Hasan, and T. Durakiewicz, Observation of topological nodal fermion semimetal phase in $\mathrm{ZrSiS}$,  \href{https://doi.org/10.1103/PhysRevB.93.201104}{Phys. Rev. B \textbf{93,} 201104(R) (2016)}.

\bibitem{bradlyn} B. Bradlyn, J. Cano, Z. Wang, M. G. Vergniory, C. Felser, R. J. Cava, and B. A. Bernevig, Beyond Dirac and Weyl fermions: Unconventional quasiparticles in conventional crystals, \href{https://doi.org/10.1126/science.aaf5037}{Science \textbf{353,} aaf5037 (2016)}.

\bibitem{CoSi} D. S. Sanchez, I. Belopolski, T. A. Cochran, X. Xu, J.-X. Yin, G. Chang, W. Xie, K. Manna, V. Süß, C.-Y. Huang, N. Alidoust, D. Multer, S. S. Zhang, N. Shumiya, X. Wang, G.-Q. Wang, T.-R. Chang, C. Felser, S.-Y. Xu, S. Jia, H. Lin, and M. Z. Hasan, Topological chiral crystals with helicoid-arc quantum states, \href{https://doi.org/10.1038/s41586-019-1037-2}{Nature \textbf{567,} 500 (2019)}.

\bibitem{topp} A. Topp, J. M. Lippmann, A. Varykhalov, V. Duppel, B. V. Lotsch, C. R. Ast, and L. M. Schoop, Non-symmorphic band degeneracy at the Fermi level in $\mathrm{ZrSiTe}$, \href{https://doi.org/10.1088/1367-2630/aa4f65}{New J. Phys. \textbf{18,} 125014 (2016)}.

\bibitem{hu} J. Hu, Z. Tang, J. Liu, X. Liu, Y. Zhu, D. Graf, K. Myhro, S. Tran, C. N. Lau, J. Wei, and Z. Mao, Evidence of Topological Nodal-Line Fermions in $\mathrm{ZrSiSe}$ and $\mathrm{ZrSiTe}$, \href{https://doi.org/10.1103/PhysRevLett.117.016602}{Phys. Rev. Lett. \textbf{117,} 016602 (2016)}.

\bibitem{takane} D. Takane, Z. Wang, S. Souma, K. Nakayama, C. X. Trang, T. Sato, T. Takahashi, and Y. Ando, Dirac-node arc in the topological line-node semimetal $\mathrm{HfSiS}$, \href{https://doi.org/10.1103/PhysRevB.94.121108}{Phys. Rev. B \textbf{94,} 121108 (2016)}.

\bibitem{zrsix} M. M. Hosen, K. Dimitri, I. Belopolski, P. Maldonado, R. Sankar, N. Dhakal, G. Dhakal, T. Cole, P. M. Oppeneer, D. Kaczorowski, F. Chou, M. Z. Hasan, T. Durakiewicz, and M. Neupane, Tunability of the topological nodal-line semimetal phase in $\mathrm{ZrSi}X$-type materials ($X=\mathrm{S}, \mathrm{Se}, \mathrm{Te}$), \href{https://doi.org/10.1103/PhysRevB.95.161101}{Phys. Rev. B \textbf{95,} 161101 (2017)}.

\bibitem{chen} C. Chen, X. Xu, J. Jiang, S. C. Wu, Y. P. Qi, L. X. Yang, M. X. Wang, Y. Sun, N. B. M. Schröter, H. F. Yang, L. M. Schoop, Y. Y. Lv, J. Zhou, Y. B. Chen, S. H. Yao, M. H. Lu, Y. F. Chen, C. Felser, B. H. Yan, Z. K. Liu, and Y. L. Chen, Dirac line nodes and effect of spin-orbit coupling in the nonsymmorphic critical semimetals $M\mathrm{SiS}\phantom{\rule{0.16em}{0ex}}(M=\mathrm{Hf},\phantom{\rule{0.16em}{0ex}}\mathrm{Zr})$, \href{https://doi.org/10.1103/PhysRevB.95.125126}{Phys. Rev. B  {95,} 125126 (2017)}.

\bibitem{lou} R. Lou, J. Z. Ma, Q. N. Xu, B. B. Fu, L. Y. Kong, Y. G. Shi, P. Richard, H. M. Weng, Z. Fang, S. S. Sun, Q. Wang, H. C. Lei, T. Qian, H. Ding, and S. C. Wang, Emergence of topological bands on the surface of $\mathrm{ZrSnTe}$ crystal, \href{https://doi.org/10.1103/PhysRevB.93.241104}{Phys. Rev. B \textbf{93,} 241104 (2016)}.

\bibitem{hu2}  J. Hu, Y. Zhu, X. Gui, D. Graf, Z. Tang, W. Xie, and Z. Mao, Quantum oscillation evidence for a topological semimetal phase in $\mathrm{ZrSnTe}$, \href{https://doi.org/10.1103/PhysRevB.97.155101}{Phys. Rev. B \textbf{97,} 155101 (2018)}.

\bibitem{zhang} J. Zhang, M. Gao, J. Zhang, X. Wang, X. Zhang, M. Zhang, W. Niu, R. Zhang, and Y. Xu, Transport evidence of 3D topological nodal-line semimetal phase in $\mathrm{ZrSiS}$, \href{https://doi.org/10.1007/s11467-017-0705-7}{Front. Phys. \textbf{13,} 137201 (2017)}.

\bibitem{zrgete} M. M. Hosen, K. Dimitri, A. Aperis, P. Maldonado, I. Belopolski, G. Dhakal, F. Kabir, C. Sims, M. Z. Hasan, D. Kaczorowski, T. Durakiewicz, P. M. Oppeneer, and M. Neupane, Observation of gapless Dirac surface states in $\mathrm{ZrGeTe}$, \href{https://doi.org/10.1103/PhysRevB.97.121103}{Phys. Rev. B \textbf{97,} 121103 (2018)}.

%\bibitem{zrgete2} Y. Yen, C.-L. Chiu, P.-H. Lin, R. Sankar, F. Chou, T.-M. Chuang, and G.-Y. Guo, Dirac Nodal Line and Rashba Splitting Surface States in Nonsymmorphic $\mathrm{ZrGeTe}$, \href{https://arxiv.org/abs/1912.07002}{arXiv:1912.07002 (2019)}.

%\bibitem{cheng} Z. Cheng, Z. Zhang, H. Sun, S. Li, H. Yuan, Z. Wang, Y. Cao, Z. Shao, Q. Bian, X. Zhang, F. Li, J. Feng, S. Ding, Z. Mao, and M. Pan, Visualizing Dirac nodal-line band structure of topological semimetal $\mathrm{ZrGeSe}$ by ARPES, \href{https://doi.org/10.1063/1.5084090}{APL Mater. \textbf{7,} 051105 (2019)}.

\bibitem{fu} B. B. Fu, C. J. Yi, T. T. Zhang, M. Caputo, J. Z. Ma, X. Gao, B. Q. Lv, L. Y. Kong, Y. B. Huang, P. Richard, M. Shi, V. N. Strocov, C. Fang, H. M. Weng, Y. G. Shi, T. Qian, and H. Ding, Dirac nodal surfaces and nodal lines in $\mathrm{ZrSiS}$, \href{https://doi.org/10.1126/sciadv.aau6459}{Sci. Adv. \textbf{5,} eaau6459 (2019)}.

\bibitem{ali} M. N. Ali, L. M. Schoop, C. Garg, J. M. Lippmann, E. Lara, B. Lotsch, and S. S. P. Parkin, Butterfly magnetoresistance, quasi-2D Dirac Fermi surface and topological phase transition in $\mathrm{ZrSiS}$, \href{https://doi.org/10.1126/sciadv.1601742}{Sci. Adv. \textbf{2,} e1601742 (2016)}.

\bibitem{lv} Y.-Y. Lv, B.-B. Zhang, X. Li, S.-H. Yao, Y. B. Chen, J. Zhou, S.-T. Zhang, M.-H. Lu, and Y.-F. Chen, Extremely large and significantly anisotropic magnetoresistance in $\mathrm{ZrSiS}$ single crystals, \href{https://doi.org/10.1063/1.4953772}{Appl. Phys. Lett. \textbf{108,} 244101 (2016)}.

\bibitem{singha} R. Singha, A. K. Pariari, B. Satpati, and P. Mandal, Large nonsaturating magnetoresistance and signature of nondegenerate Dirac nodes in $\mathrm{ZrSiS}$, \href{https://doi.org/10.1073/pnas.1618004114}{Proc. Natl. Acad. Sci. \textbf{114,} 2468 (2017)}.

\bibitem{kumar} N. Kumar, K. Manna, Y. Qi, S.-C. Wu, L. Wang, B. Yan, C. Felser, and C. Shekhar, Unusual magnetotransport from Si-square nets in topological semimetal $\mathrm{HfSiS}$, \href{https://doi.org/10.1103/PhysRevB.95.121109}{Phys. Rev. B \textbf{95,} 121109 (2017)}.

\bibitem{schilling} M. B. Schilling, L. M. Schoop, B. V. Lotsch, M. Dressel, and A. V. Pronin, Flat Optical Conductivity in $\mathrm{ZrSiS}$ due to Two-Dimensional Dirac Bands, \href{https://doi.org/10.1103/PhysRevLett.119.187401}{Phys. Rev. Lett. \textbf{119,} 187401 (2017)}.

\bibitem{pezzini}  S. Pezzini, M. R. van Delft, L. M. Schoop, B. V. Lotsch, A. Carrington, M. I. Katsnelson, N. E. Hussey, and S. Wiedmann, Unconventional mass enhancement around the Dirac nodal loop in $\mathrm{ZrSiS}$, \href{https://doi.org/10.1038/nphys4306}{Nat. Phys. \textbf{14,} 178 (2018)}.

\bibitem{gdsbte} M. M. Hosen, G. Dhakal, K. Dimitri, P. Maldonado, A. Aperis, F. Kabir, C. Sims, P. Riseborough, P. M. Oppeneer, D. Kaczorowski, T. Durakiewicz, and M. Neupane, Discovery of topological nodal-line fermionic phase in a magnetic material $\mathrm{GdSbTe}$, \href{https://doi.org/10.1038/s41598-018-31296-7}{Sci. Rep. \textbf{8,} 13283 (2018)}.

\bibitem{cesbte} L. M. Schoop, A. Topp, J. Lippmann, F. Orlandi, L. Müchler, M. G. Vergniory, Y. Sun, A. W. Rost, V. Duppel, M. Krivenkov, S. Sheoran, P. Manuel, A. Varykhalov, B. Yan, R. K. Kremer, C. R. Ast, and B. V. Lotsch, Tunable Weyl and Dirac states in the nonsymmorphic compound $\mathrm{CeSbTe}$, \href{https://doi.org/10.1126/sciadv.aar2317}{Sci. Adv. \textbf{4,} eaar2317 (2018)}.

\bibitem{cesbte2} A. Topp, M. G. Vergniory, M. Krivenkov, A. Varykhalov, F. Rodolakis, J. L. McChesney, B. V. Lotsch, C. R. Ast, and L. M. Schoop, The effect of spin-orbit coupling on nonsymmorphic square-net compounds, \href{https://doi.org/10.1016/j.jpcs.2017.12.035}{Journal of Physics and Chemistry of Solids \textbf{128,} 296 (2019)}.
  
 \bibitem{hosbte} S. Yue, Y. Qian, M. Yang, D. Geng, C. Yi, S. Kumar, K. Shimada, P. Cheng, L. Chen, Z. Wang, H. Weng, Y. Shi, K. Wu, and B. Feng, Topological electronic structure in the antiferromagnet $\mathrm{HoSbTe}$, \href{https://doi.org/10.1103/PhysRevB.102.155109}{Phys. Rev. B \textbf{102,} 155109 (2020)}.
 
 \bibitem{ndsbte} K. Pandey, R. Basnet, A. Wegner, G. Acharya, M. R. U. Nabi, J. Liu, J. Wang, Y. K. Takahashi, B. Da, and J. Hu, Electronic and magnetic properties of the topological semimetal candidate $\mathrm{NdSbTe}$, \href{https://doi.org/10.1103/PhysRevB.101.235161}{Phys. Rev. B \textbf{101,} 235161 (2020)}.
 
 \bibitem{cesbte3} B. Lv, J. Chen, L. Qiao, J. Ma, X. Yang, M. Li, M. Wang, Q. Tao, and Z.-A. Xu, Magnetic and transport properties of low-carrier-density Kondo semimetal $\mathrm{CeSbTe}$, \href{https://doi.org/10.1088/1361-648X/ab2498}{J. Phys. Condens. Mater. \textbf{31,} 355601 (2019)}.
 
 \bibitem{lasbte} Y. Wang, Y. Qian, M. Yang, H. Chen, C. Li, Z. Tan, Y. Cai, W. Zhao, S. Gao, Y. Feng, S. Kumar, E. F. Schwier, L. Zhao, H. Weng, Y. Shi, G. Wang, Y. Song, Y. Huang, K. Shimada, Z. Xu, X. J. Zhou, and G. Liu, Spectroscopic evidence for the realization of a genuine topological nodal-line semimetal in $\mathrm{LaSbTe}$, \href{https://doi.org/10.1103/PhysRevB.103.125131}{Phys. Rev. B \textbf{103,} 125131 (2021)}.
 
\bibitem{SI} For more information, see the supplemental material to this paper, which includes Refs. \cite{Hohe64, Kohn65, Bloo94, Kres96, Kres96.1, Perd96, Duda98, Monk76, Lope85, Most14, Wu18}. 
 
\bibitem{Hohe64} P. Hohenberg, W. Kohn,  Inhomogeneous Electron Gas,  \href{https://doi.org/10.1103/PhysRev.136.B864}{Phys. Rev. \textbf{136,} B864 (1964)}.

\bibitem{Kohn65} W. Kohn, L. J. Sham, Self-Consistent Equations Including Exchange and Correlation Effects, \href{https://doi.org/10.1103/PhysRev.140.A1133}{Phys. Rev. \textbf{140,} A1133 (1965)}.

\bibitem{Bloo94} P. E. Blöchl, Projector augmented-wave method,  \href{https://doi.org/10.1103/PhysRevB.50.17953}{Phys. Rev. B \textbf{50,} 17953 (1994)}.

\bibitem{Kres96} G. Kresse and J. Furthmüller, Efficient iterative schemes for ab initio total-energy calculations using a plane-wave basis set, \href{https://doi.org/10.1103/PhysRevB.54.11169}{Phys. Rev. B \textbf{54,} 11169 (1996)}.

\bibitem{Kres96.1} G. Kresse and J. Furthmüller, Efficiency of ab-initio total energy calculations for metals and semiconductors using a plane-wave basis set, \href{https://doi.org/10.1016/0927-0256(96)00008-0}{Comp. Mater. Sci. \textbf{6,} 15 (1996)}.
 
 \bibitem{Ando1} Z. Ren, A. A. Taskin, S. Sasaki, K. Segawa, and Y. Ando, Large bulk resistivity and surface quantum oscillations in the topological insulator ${\text{Bi}}_{2}{\text{Te}}_{2}\text{Se}$, \href{https://doi.org/10.1103/PhysRevB.82.241306}{Phys. Rev. B \textbf{82,} 241306(R)(2010)}.

\bibitem{Ando2} A. A. Taskin, Z. Ren, S. Sasaki, K. Segawa, and Y. Ando, Observation of Dirac Holes and Electrons in a Topological Insulator, \href{https://doi.org/10.1103/PhysRevLett.107.016801}{Phys. Rev. Lett. \textbf{107,} 016801 (2011)}.

\bibitem{Perd96} J. P. Perdew, K. Burke, and M. Ernzerhof, Generalized Gradient Approximation Made Simple, \href{https://doi.org/10.1103/PhysRevLett.77.3865}{Phys. Rev. Lett. \textbf{77}, 3865 (1996)}.

\bibitem{Duda98} S. L. Dudarev, G. A. Botton, S. Y. Savrasov, C. J. Humphreys, and A. P. Sutton, Electron-energy-loss spectra and the structural stability of nickel oxide: An LSDA+U study, \href{https://doi.org/10.1103/PhysRevB.57.1505}{Phys. Rev. B \textbf{57}, 1505 (1998)}.

\bibitem{Monk76}H. J. Monkhorst and J. D. Pack, Special points for Brillouin-zone integrations, \href{https://doi.org/10.1103/PhysRevB.13.5188}{Phys. Rev. B \textbf{13}, 5188 (1976)}.

\bibitem{Lope85} M. P. L. Sancho, J. M. L. Sancho, J. M. L. Sancho, and J. Rubio, Highly convergent schemes for the calculation of bulk and surface Green functions,  \href{https://doi.org/10.1088/0305-4608/15/4/009}{J. Phys. F: Met. Phys. \textbf{15}, 851 (1985)}.

\bibitem{Most14} A. A. Mostofi, J. R. Yates, G. Pizzi, Y.-S. Lee, I. Souza, D. Vanderbilt, and N. Marzari, An updated version of wannier90: A tool for obtaining maximally-localised Wannier functions, \href{https://doi.org/10.1016/j.cpc.2014.05.003}{Comput. Phys. Commun. \textbf{185}, 2309 (2014)}.

\bibitem{Wu18} Q. Wu, S. Zhang, H.-F. Song, M. Troyer, and A. A. Soluyanov, WannierTools: An open-source software package for novel topological materials,  \href{https://doi.org/10.1016/j.cpc.2017.09.033}{Comput. Phys. Commun. \textbf{224}, 405 (2018)}.
\end{thebibliography}
\end{document}